\documentclass[%
 reprint,
 amsmath,amssymb,
 aps,
]{revtex4-2}

\usepackage{graphicx}
\usepackage{dcolumn}
\usepackage{bm}

\begin{document}

\preprint{APS/123-QED}

\title{Longitudinal spin-relaxation optimization for miniaturized optically pumped magnetometers}

\author{A. P. McWilliam}
\email{allan.mcwilliam@strath.ac.uk}
\author{S. Dyer}
\author{D. Hunter}
\author{M. Mrozowski}
\author{S. J. Ingleby}
\author{O. Sharp}
\author{D. P. Burt}
\author{P. F. Griffin}
\author{J. P. McGilligan}
\author{E. Riis}

\affiliation{Department of Physics, SUPA, University of Strathclyde, Glasgow G4 0NG, United Kingdom}
\affiliation{\textsuperscript{2}Kelvin Nanotechnology, University of Glasgow, Glasgow G12 8LS, United Kingdom}

\date{\today}





\date{\today}

\begin{abstract}
The microfabrication of cesium vapor cells for optically pumped magnetometry relies on optimization of buffer gas pressure in order to maximize atomic coherence time and sensitivity to external magnetic signals. We demonstrate post-bond nitrogen buffer gas pressure tuning through localized heating of an integrated micro-pill dispenser. We characterize the variation in the intrinsic longitudinal relaxation rate, $\gamma_{10}$, and magnetic sensitivity, as a function of the resulting nitrogen buffer gas pressure. Measurements are conducted through employing an optically pumped magnetometer operating in a free-induction-decay configuration. $\gamma_{10}$ is extracted across a range of nitrogen pressures between $\sim$~60~-~700~Torr, measuring a minimum of 140~Hz at 115~Torr. Additionally, we achieve sensitivities as low as 130 ~fT/$\sqrt{\text{Hz}}$ at a bias field amplitude of $\sim 50~$µT. With the optimal nitrogen buffer gas pressure now quantified and achievable post-fabrication, these mass-producible cells can be tailored to suit a variety of sensing applications, ensuring peak magnetometer performance.

\end{abstract}

\maketitle


\section{\label{sec:level1}Introduction}

Precise magnetic field sensing in finite field conditions is crucial in several areas including geophysical surveying \cite{geosurvey}, land ordinance \cite{ordinance} and archaeology \cite{archae}. Additionally, continuous research has revealed promising applications in the medical field including magnetocardiography (MCG) and magnetoencephalography (MEG) \cite{twinleaf2020}. These applications typically require a network of sensors to adequately resolve both spatial and temporal information, and often necessitate gradiometric cancellation when operated in unshielded conditions \cite{unshielded_MEG_Grad}. Therefore, precise tuning of the buffer gas pressure offers significant advantages by providing consistent sensor performance, improved common-mode noise suppression, and a uniform response across the sensor array.   

Optically pumped magnetometers (OPMs) have emerged as a versatile solution for satisfying the numerous sensing requirements across these avenues. OPMs measure the precession frequency of spin-polarized atoms, providing extremely accurate and precise magnetic field readings \cite{2007budker}. Importantly, the incorporation of thermal atomic ensembles in OPMs has enabled miniaturization into portable and chip-scale sensor heads to address real-world implementations \cite{ingleby2022digital}.

At the heart of the OPM comprises an atomic vapor cell, where atom-light interactions occur. Such vapor cells contain the atomic spin medium, in conjunction with a buffer gas or paraffin coating, required to extend the atomic spin-coherence time. A core component in the miniaturization of OPMs is the micro-machined vapor cell, typically formed of a glass-silicon-glass stack containing buffer gas to slow the diffusion of the alkali atoms towards the cell walls. Such cells are commonly manufactured with a uniform target buffer gas pressure intended across the entire wafer \cite{mcgilligan_review}. The reproducibility and achievable yield is on the order of 100's of cells per wafer making this approach advantageous for development of sensor arrays and mass-producible quantum technologies \cite{li2024waferscale}, with custom geometries and multi-chamber designs realizable \cite{s_dyer_2022_multi_chamber,raghavan2024functionalized,kitching_chipscale_review}. Additionally, precise buffer gas control enables the optimization between spatial resolution and sensor precision in OPM magnetic imaging applications \cite{hunter2023imaging}. This refinement enables an optimal balance between two factors for peak performance: spin-destruction collisions between alkali and buffer gas atoms, and the atomic diffusion rate. 

We employ an OPM based on the free-induction-decay (FID) measurement protocol \cite{Fiddom2018}. One significant advantage of FID-based sensors is their insensitivity to systematics, achieved by temporally separating the optical pumping and detection processes using a pulsed measurement scheme. This separation allows the polarized spins to precess freely at the Larmor frequency without interference from intense pumping light, significantly reducing light shifts and power broadening effects compared to continuous-wave pumping schemes. Additionally, an enhanced optical pumping mode is exploited to facilitate suppression of spin-exchange collisions, reducing the overall spin-relaxation contribution \cite{DOM_ESP_paper_2023}. This vastly extends the sensor's dynamic range to bias fields surpassing the Earth's field ($\sim~50~$µT).

In this paper, we demonstrate the ability to optimize the longitudinal spin-relaxation rate in a micro-machined vapor cell by controllable gettering of the nitrogen (N$_2$) buffer gas pressure from localized heating of an alkali micro-pill source. The intrinsic spin-relaxation properties for a range of vapor cells is ascertained using the FID approach \cite{Fiddom2018, waveformdom, DOM_ESP_paper_2023, hunter2023imaging}. Power broadening caused by residual optical pumping from the probe light is mitigated through extrapolation to zero-light power \cite{scholtes2014intrinsic}. Furthermore, the impact of the buffer gas pressure on the measured sensitivity of the OPM is shown for cell pressures measured between $60-700$~Torr. The ability to optimize the OPM sensitivity and longitudinal relaxation rate by reducing the vapor cell buffer gas pressure in a single vapor cell gives critical insight to the operational characteristics of the sensor. Our findings will guide future mass production of cells with the optimal N$_2$ pressure content to be applied via back-filling now quantified. Moreover, the N$_2$ content can be equalized across a population of cells achieving uniform response in magnetic gradiometers and OPM sensor arrays.

\section{Methodology}
\subsection{Experimental setup}
\label{sec:FId_exp_setup}

\begin{figure*}[t!]
\centering
\includegraphics[width=0.8\linewidth]{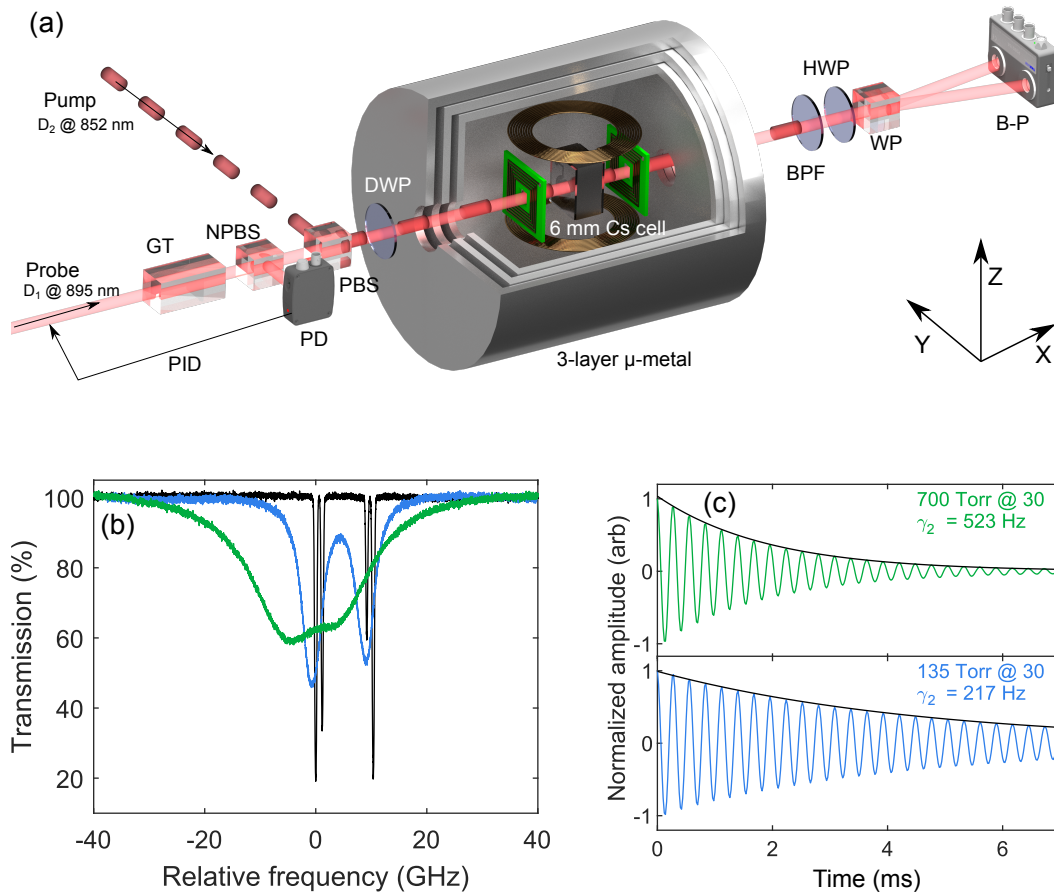}
    \caption{(a) OPM experimental setup containing 852~nm and 895~nm pump and probe beams that interact with cesium (Cs) atoms in a thermal vapor cell placed in a magnetic shield with a controlled bias field applied along the z-axis. The transmitted probe was detected with a balanced polarimeter while the residual pump was blocked with a spectral filter. GT: Glan-Thompson polarizer; NPBS: Non-polarizing beam-splitter; PBS: Polarizing beam-splitter; DWP: Dual-wavelength waveplate; BPF: Band-pass filter; WP: Wollaston prism; B-P: Balanced polarimeter; PD: Photodiode used in conjunction with an AOM (not pictured) for probe beam intensity stabilization. (b) Absorption spectroscopy (Cs $\text{D}_1$ line) was used to extract (N$_2$) pressure readings with spectra shown for 700 Torr (green) and 135 Torr (blue) compared against a reference cell (black). (c) Corresponding FID signal data relating to a probe power of 1.3$~$mW along with relaxation envelopes (black). The probe beam induces a broadening effect, $\gamma_\text{P}$, leading to an increase in the relaxation rate.}
    \label{fig:schematic}
\end{figure*}

 Figure \ref{fig:schematic}(a) displays a simplified schematic of the experimental setup used in this work. We employed three separate vapor cells, which were anodically bonded with different N$_2$ pressures (P$_{\text{N}_2}$). Each cell had dimensions of ($6~\times~6~\times~6$)~mm within the science chamber. Two cells were cylindrically shaped whilst the third was a cuboid geometry. We conducted sequential measurements for each using this OPM setup. The N$_2$ content was decreased by laser heating the alkali pill source to activate the non-evaporable getter material inside as discussed in \cite{dyer2023realtime, lucivero2022neg}. P$_{\text{N}_2}$ within the cells was determined after each depletion step by performing absorption spectroscopy on the Cs $\text{D}_1$ line with the cell heated to 70 $^\circ$C. The resulting spectrum was then fitted with an absorption profile to extract the N$_2$ induced pressure broadening and shift, which were used in conjunction with the coefficients characterized by Andalkar \textit{et al.} \cite{Andalkar2002pressureshiftbroadening} to determine the P$_{\text{N}_2}$. P$_{\text{N}_2}$ scales as a function of cell temperature according to the ideal gas law. The cell was operated at 30 $^\circ$C for spin-relaxation measurements to suppress the spin-exchange contributions observed at higher vapor densities, although an appreciable signal could still be observed. The sensitivity was measured at 70 $^\circ$C which is optimal for the vapor cell configurations considered here. For each measurement throughout this study the cell was placed within a three-layer µ-metal shield to attenuate the Earth's ambient field amplitude and sources of environmental noise. \\
 \indent A set of coils were thermally connected to both sides of the cell using printed circuit boards (PCBs), as depicted in Fig. \ref{fig:schematic}(a). These coils provided resistive heating and established a well-defined optical pumping (quantization) axis, ensuring high spin-polarization even in the presence of substantial bias fields. This is due to the strong magnetic field, $|\vec{\text{B}}_{\text{Pol}}|>1~$mT, applied along the quantization axis by the coils, which greatly exceeds the fields within the measurement range of interest \cite{DOM_ESP_paper_2023}. Surrounding the cell were a set of Helmholtz coils which provided static transverse magnetic field control. This coil set was driven by a programmable low-noise current driver that does not contribute significantly to the overall noise budget of the sensor \cite{marcinHCCD}. \\
 \indent We operated a two-color pump-probe scheme as described in previous works \cite{hunter2023imaging, DOM_ESP_paper_2023}. Strong pulsed optical pumping was applied ($\sim 120$~mW peak power) with circularly polarized light on the pressure broadened \textit{F}~=~3$~\rightarrow$~\textit{F}' hyperfine transition of the Cs $\text{D}_2$ line. The pump light was extinguished from peak optical power to $<$~5~µW using an acousto-optic modulator (AOM). The pump light recycles atomic population present in the \textit{F}~=~3 manifold which are subsequently redistributed to the \textit{F}~=~4 ground state, inducing an orientation in the atomic ensemble. This is akin to the hyperfine-repump technique employed in \cite{2015_lightnarrowSchultze:15}, albeit without an additional laser source. Moreover, the continuous absorption and emission process transfers a large net spin-polarization into a stretched Zeeman state with maximum angular momentum $m_F=4$. Collision of atoms occupying the same hyperfine ground state do not result in decoherence through spin-exchange, therefore efficient optical pumping to a highly stretched state allows effective suppression of spin-exchange dephasing \cite{2011_lightnarrow}. 

Detection was performed on the Cs $\text{D}_1$ line with a weaker probe beam, less than 1.2~\% of the pump beam's peak optical power. The probe frequency was set $\approx$~21~GHz blue-detuned from the \textit{F}~=~3$~\rightarrow$~\textit{F'} transition. This was close enough to resonance to yield an appreciable signal amplitude, whilst avoiding excessive power broadening due to residual optical pumping. To obtain the intrinsic spin-relaxation property, $\gamma_{10}$, it was necessary to negate these operational systematics further. This was achieved by applying a range of probe powers from $\sim$ 0.1~mW to 1.3~mW such that $\gamma_{10}$ could be ascertained by extrapolating to zero-light power. The probe power was measured before transmission through the cell which was $\geq$~70~\%. The pump and probe beams were ensured to overlap at the center of the cell, where they both possessed a 3.1~mm beam diameter ($1/\text{e}^2$). Therefore, both were sufficiently well contained inside the cell volume to avoid beam clipping.  
 
\subsection{Signal analysis}
\label{sec:fitting}

A balanced polarimeter (Thorlabs PDB210A), which rejects common-mode laser intensity noise, was used to capture the signal. A data acquisition device (Picoscope Model 5444D) sampled at a base rate of f$_\text{s}$ = 125~MHz and functioned with 15-bit voltage resolution. Over-sampling was performed to average 25 successive data points resulting in a decimated sample rate of $\text{f}_\text{s}=1/\Delta\text{t}=$~5~MHz. This was due to f$_\text{s}$ greatly exceeding the signals of interest, which oscillated at either 3.5~kHz (Section \ref{Sec:relaxrate_results}) or 175~kHz (Section \ref{sec:sens_results}). Furthermore, for the former in post-process a low pass Butterworth filter (2nd order) was applied to suppress any spurious higher frequency noise. This data was also further down-sampled by a factor of  100 to reduce the effective sample rate to f$_\text{s}$ = 50 kHz (equating to approximately 14 points per cycle). This served only to expedite the fitting analysis and did not corrupt the extracted fit parameters.

The FID signal manifests as a result of optical rotation of the probe beam polarization, which maps the induced spin-polarization precessing around the magnetic field at the Larmor frequency. The digitized FID signal can be represented as follows, 
\begin{equation}
    \text{S}_\text{n} = \text{A}\,\mathrm{sin}\big(\omega_{\text{L}}\,\text{n}\,\Delta{\text{t}} + \phi_{0}\big)\,e^{-\gamma_2\,\text{n}\,\Delta{\text{t}}} + \epsilon_\text{n}.
    \label{Eq:Model-fit}
\end{equation}

\noindent
where \text{A} is the signal amplitude, relating to the spin-polarization generated during the optical pumping phase, $\omega_{\text{L}}$ is the Larmor precession frequency, and $\phi_0$ is the initial phase. \text{n} relates to the sample data point under consideration and $\epsilon_\text{n}$ represents noise present in the signal. Nonlinear fitting was applied with the Levenberg-Marquardt algorithm \cite{hughes2010measurements} using a damped sinusoidal model to extract the relevant parameters. After optical pumping, the precessing atomic spin-polarization decays as the polarization returns to its equilibrium (unpolarized) state, with a total spin-relaxation rate denoted $\gamma_2$. Section \ref{sec:Relax rate motivation} describes methods of lowering $\gamma_2$ such that this equates to the intrinsic longitudinal rate of the vapor cell. This was enabled by reducing the cell temperature and probe power with the results presented in Section \ref{Sec:relaxrate_results}. In Section \ref{sec:sens_results} our chosen operating regime involved the vapor density being raised to increase atom-light interactions, thus elevating $\gamma_2$. However, this ultimately enhanced the sensitivity and provided a better analog for a practical sensor with optimized performance.



\subsection{Relaxation rate measurements}
\label{sec:Relax rate motivation}

The spin-relaxation rate limits the total possible measurement time of the OPM. The various intrinsic relaxation mechanisms inherent to vapor cells are extensively described in \cite{IRR_2014}. The intrinsic longitudinal relaxation rate, $\gamma_{10}$, is the ultimate spin-relaxation limit. The main contributions to this in this study arise from alkali-wall collisions ($\gamma_{\text{WC}}$) and alkali-buffer gas collisions ($\gamma_{\text{BG}}$). A negligible contribution to the spin-decoherence arises through Cs-Cs spin-destruction collisions ($\gamma_{\text{SD}}$). The cross section for this is a factor of $\sim$~100 smaller than Cs-Cs spin-exchange collisions \cite{Spin_exchange_Ressler, Spin_destruction_Walker}. Therefore, $\gamma_{10}$ is defined as,
\begin{equation}
    \gamma_{10} = \gamma_{\text{WC}} + \gamma_{\text{BG}} + \gamma_{\text{SD}}.
    \label{eq:theory_gamma1}
\end{equation}
The aggregated term representing the total transverse relaxation rate, $\gamma_2$, including operational systematics, is given by,
\begin{equation}
    \gamma_{\text{2}} = \gamma_{10} + \gamma_{\text{SE}} + \gamma_\text{P} + \gamma_{\Delta\text{B}}.
    \label{eq:theory_gamma1}
\end{equation}
Therefore, to successfully extract the intrinsic relaxation rate, $\gamma_{10}$, the additional relaxation contributions including optical power broadening ($\gamma_\text{P}$), magnetic gradients across the cell ($\gamma_{\Delta\text{B}}$), and spin-exchange collisions ($\gamma_{\text{SE}}$) have to be suppressed. 

Operating the system at a low bias magnetic field strength of $|\vec{\text{B}_\text{z}}| \approx$~1~µT lowered magnetic gradients to $\gamma_{\Delta\text{B}}<1$~Hz based on calculations considering the coil geometry and vapor cell dimensions. Furthermore, spin-exchange collisions are negligible at the low cell temperatures (T~=~30~$^\circ$C) considered here. Temperature was monitored using three non-magnetic T-type thermocouple sensors attached to separate positions of the cell. All agreed to within 0.1~$^{\circ}$C, confirming a uniform temperature across the cell. The cell was not found to drift by more than 0.1~$^{\circ}$C over the course of each set of measurements. It should be noted that an insignificant increase in $\gamma_{10}$ occurs for increasing temperatures.

The pump-probe cycle was applied at a repetition rate, $\text{R}_\text{r}$, of 10~Hz with a pump duration, $\text{T}_\text{OP}$, of 5~ms and total probing duration, $\text{T}_\text{PR}$, of 95~ms. This measurement period ensured the atoms fully decohered and re-thermalized before subsequent optical pumping cycles occurred, thus avoiding fitting systematics from signal truncation. To determine the intrinsic relaxation rate under these conditions, measurements were taken at various probe powers, starting from a maximum of 1.3$~$mW and decreasing incrementally to 140~µW. Nonlinear fitting using a damped sinusoidal model (Eq. \ref{Eq:Model-fit}) was applied to the FID datasets to extract $\gamma_\text{2}$ in each case. A linear extrapolation to zero-light power accounts for the contribution from $\gamma_\text{P}$ such that $\gamma_\text{10}$ can be determined as seen in Fig. \ref{fig:relaxresults} (a).

\section{Results}
\subsection{Relaxation rate results}
\label{Sec:relaxrate_results}

\begin{figure}[h]
    \centering
    \includegraphics{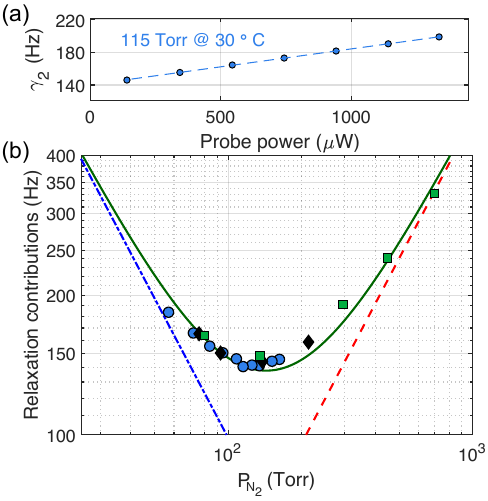}
    \caption{(a) Measured values for $\gamma_{\text{2}}$ as a function of probe power for P$_{\text{N}_2}=115$~Torr, where $\gamma_{10}$ is minimized. Extrapolation to zero-light power at each N$_2$ pressure provides the respective $\gamma_{10}$ values shown in (b). The spin-exchange and magnetic field broadening contributions to $\gamma_{\text{2}}$ were minimal throughout these spin-relaxation measurements. (b) Blue data points relate to measurements with the cuboid cell. Black (green) data points refer to two separate cylindrical shaped cells. The various lines relate to the theoretical spin-relaxation mechanisms: $\gamma_{10}$ (green solid line); $\gamma_{\text{WC}}$ (blue dot-dashed line); $\gamma_{\text{BG}}$ (red dashed line). Error-bars are smaller than the markers.}

    
    \label{fig:relaxresults}
\end{figure}

Figure \ref{fig:relaxresults} (b) displays the results over the full buffer gas pressure range.  Each pressure iteration produced small statistical errors for each data point on the order of 1~Hz. The data demonstrates a continuous optimization of $\gamma_{10}$ when lowering $\text{P}_{\text{N}_2}$ to 115~Torr, which is where it is minimized. $\gamma_{10}$ then begins to increase as $\text{P}_{\text{N}_2}$ is reduced further due to Cs-wall collisions ($\gamma_{\text{WC}}$) becoming the dominant depolarizing mechanism and the contribution from Cs-$\text{N}_2$ collisions ($\gamma_{\text{BG}}$) diminishing. Across the pressure range $\text{P}_{\text{N}_2} \approx 57-212$~Torr, the intrinsic relaxation rate extends from $\gamma_{10} \approx 140-184$~Hz. This equates to intrinsic spin-coherence times of between 5.4~ms and 7.1~ms. Moreover, $\gamma_{10}$ maintains a consistency below $150$~Hz between a range of $108-162$~Torr, showcasing a fairly wide pressure range that permits extensive coherence times. This is a significant time period in which magnetic sensing measurements can be made before spin-relaxation occurs providing enhanced sensitivity performance. \\
\indent The data aligns with the theoretical model outlined in \cite{IRR_2014} when constraining all known parameters, with the exception of two critical parameters; namely, the Cs-N$_2$ spin-destruction rate, $\sigma_{\text{Cs}-\text{N}_2}$, and the Cs diffusion coefficient in N$_2$, $\text{D}_{0:\text{Cs}-\text{N}_2}$. The data indicates that these values are $\sigma_{\text{Cs}-\text{N}_2}$ = 2.9 x 10$^{-26}$ m$^2$ and  $\text{D}_{0:\text{Cs}-\text{N}_2}$ = 0.12 $\text{cm}^2\text{s}^{-1}$ respectively. This value for $\text{D}_{0:\text{Cs}-\text{N}_2}$ matches well with the works from \cite{Franz1964,beverini_strumia_1971, Franz1974upgrade, kitchingknappehollberg2002} which have published values for each. $\sigma_{\text{Cs}-\text{N}_2}$ resides within a fairly wide spectrum of these published results. More recent data for these parameters are not readily available. The disparity of measured values from the literature makes it reasonable to incorporate these values in the model as they concur with the observed data. This alignment is further supported by the reliability of our measurements, involving multiple data points with three distinct cells which agree for similar $\text{P}_{\text{N}_2}$ values. \\
\indent Previously, we have demonstrated that our heating and enhanced optical pumping strategy does not cause additional systematics to $\gamma_{10}$ by perturbing the precessing ensemble \cite{DOM_ESP_paper_2023}. This was further verified in this work by collecting data at room temperature (T~$\approx 21$~$^{\circ}$C), without current being applied through the resistive heating coils. We obtained a value of $\gamma_{10} \approx 139$~Hz at $\text{P}_{\text{N}_2}\approx125$~Torr which is 2~Hz lower than the 141~Hz extracted when the cell was heated to 30~$^{\circ}$C. This agrees with theoretical predictions for $\Delta\gamma_{10}$ for this temperature variation.

\subsection{Sensitivity results}
\label{sec:sens_results}

\begin{figure}[h]
    \centering
    \includegraphics{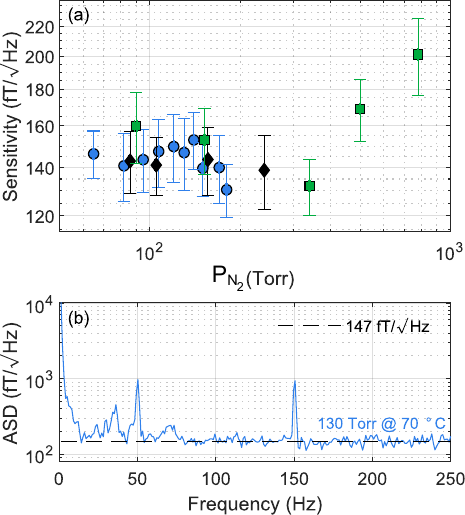}
    \caption{(a) OPM sensitivity measured across a range of N$_2$ pressures at $|\vec{\text{B}_\text{z}}| \sim50~$µT bias field. The error-bars were calculated from the standard deviation in the noise densities from each associated ASD over a 70 - 250 Hz frequency range, excluding technical noise peaks. Negligible change occurs between $\mathrm{P_{N_2}} \approx 57 - 340$~Torr providing an extensive buffer gas target range to aim for in future fabrication for optimal sensitivity performance. (b) ASD at $\mathrm{P_{N_2}} \approx 130$~Torr, $\text{T} \approx 70~^{\circ}\text{C}$, and a probe power of 1.2~mW. A sensitivity of 147~$\pm$~11~fT$/\sqrt{\text{Hz}}$ was calculated.}
    \label{fig:sensitivities}
\end{figure}

Assessment of the achievable sensitivity performance of the sensor required raising of the vapor density. Therefore, the cell was heated to 70~$^{\circ}$C for each $\mathrm{P_{N_2}}$ iteration. This elevated temperature results in an increase in the rate of spin-exchange collisions. Demonstrating the system's candidacy for unshielded applications also required increasing the bias field to $|\vec{\text{B}_\text{z}}| \approx 50$~µT. To minimize technical magnetic noise contributions, e.g. power line noise, the current through the Helmholtz coils was driven by connecting a 12~V battery to a resistor in series. Consistent pump and probe powers of 120~mW and 1.2~mW were respectively employed throughout. $\text{R}_\text{r}$, was set to 500~Hz with an optical pumping duty cycle of 10~\%, equating to $\text{T}_\text{OP}$ = 0.2~ms and $\text{T}_\text{PR}$ = 1.8~ms for each FID cycle. A faster $\text{R}_\text{r}$ here aids the sensitivity performance, which can be optimized based on the decoherence rate and the CRLB condition \cite{2010gemmelknappecrlb, Jaufenthaler_crlb}. In contrast to Section \ref{Sec:relaxrate_results}, subsequent optical pumping cycles were commenced before the atoms fully decohered or re-thermalized to maximize sensor bandwidth and sensitivity. \\
\indent Each evaluation of the magnetic sensitivity involved capturing a sequence of FID signal trains. 40 independent FID trains were recorded over 1~s time periods and subsequently processed to provide a time series of the magnetic field magnitude. The corresponding amplitude spectral densities (ASDs) were averaged using Welch's method to provide a smoother spectrum \cite{welch}. Figure \ref{fig:sensitivities} (a) displays the sensitivity results across the full buffer gas range for each cell. An example of the ASD for $\mathrm{P_{N_2}}$ = 130 Torr is displayed in (b). \\
\indent The experimental sensitivities improved as $\mathrm{P_{N_2}}$ was reduced from 780~-~340~Torr. Between 57~-~340~Torr a consistent performance was found with measured experimental sensitivities between 130~-~160~fT$/\sqrt{\text{Hz}}$. The best sensitivity of 130$~\pm~$11~fT$/\sqrt{\text{Hz}}$ was found to occur at a buffer gas pressure of 180~Torr. Notably, line and technical noise peaks are observable, indicating coupling through the electronics and magnetic environment. Variations in the amplitude and frequencies of these peaks are also evident. \\
\indent These results equate to around 3~ppb in fractional sensitivity terms \cite{wilsonperrella_frac} representing a state-of-the-art resolution using a micro-machined vapor cell. These results owe much to the manufacturing, with the 6~mm thickness reducing the relaxation rates whilst also improving the signal amplitude in comparison to our previous works with 3~mm thick cells \cite{hunter2023imaging, DOM_ESP_paper_2023}. Additionally, the intense pump light combined with applying $\vec{\text{B}}_{\text{Pol}}$ greatly assists in maintaining the spin-polarization during the atomic state preparation. Micro-machined cells have previously shown higher sensitivities than the work here, however, this was at a lower magnetic field amplitude \cite{gerginov2020scalar}. In that work, magnetic field nulling was used during the optical pumping stage, ultimately producing a sensitivity of around 100~fT$/\sqrt{\text{Hz}}$. While demonstrating excellent sensitivity, the advantage of our approach is that it circumvents the need for measuring the field and subsequently nulling it. This presents an advantage for unshielded sensing as it reduces dead-time and bypasses potential issues incurred in dynamic magnetic environments, e.g., sensing on board moving vehicles or vessels.

\section{Conclusion}

Our micro-machining capabilities have led to successful production of vapor cells with 6~mm optical path lengths (single pass). Extensive measurements were made to extract the intrinsic longitudinal spin-relaxation rate $\gamma_{10}$, which sets the ultimate limit on spin-relaxation, across a range of buffer gas pressures using three distinct vapor cells. We have demonstrated a successful application of buffer gas pressure tuning to reach the optimal value where $\gamma_{10}$ is minimized. A minimum of 140~Hz was found, measured at a cell temperature of 30~$^{\circ}$C and $\mathrm{P_{N_2}=115}$~Torr. Further reduction of $\gamma_{10}$ will be challenging without further increasing the cell size or thickness. Pivoting to fabrication approaches which do not include a buffer gas would circumvent these spin-destructive alkali-buffer gas collisions, however, the anti-relaxation coatings required to preserve spin-coherence are not compatible with the temperatures required for anodic bonding. Overall, our results indicate that the differing geometry of cells does not markedly impact the cell performance characteristics. This is an encouraging result which bodes well for mass cell fabrication as this indicates a level of consistency. A noteworthy finding is the large buffer gas pressure range at which prolonged relaxation times are achieved. \\
\indent We have demonstrated sensitivities reaching state-of-the-art levels for microfabricated vapor cells in bias fields emulating Earth's field conditions ($\sim50~$µT). We obtained a minimum of 130~$\pm$~11~fT$/\sqrt{\text{Hz}}$. Clearly sensitivities at these levels in geophysical fields is of value in several finite-field OPM applications. Additionally, our results validate the fabrication and activation approach employed for these cells. This study has demonstrated the large buffer gas pressure range at which exemplary sensor performance is attainable. One of the main advantages of the buffer gas tuning approach is the ability to decrease the pressure post-fabrication. Recent studies have demonstrated technical capabilities to sequentially increase the buffer gas pressure within microfabricated cells via a break-seal technique that well compliments the methods demonstrated in this work \cite{Maurice2022}. Magnetic gradiometers would undoubtedly benefit from this, as the capacity to finely adjust the pressure to achieve uniformity across identical cell geometries is desirable \cite{2020portable, Sheng_2017}. \\

\section*{acknowledgments}
APM was supported by a Ph.D. studentship from the Defence Science and Technology Laboratory (DSTL). JPM gratefully acknowledges funding from a Royal Academy of Engineering Research Fellowship.

\section*{Data Availability Statement}
The data that support the findings of
this study are available from the
corresponding author upon reasonable
request.

\appendix



\bibliography{apssamp}

\end{document}